\documentclass[trackchanges,twocolumn]{aastex701}

\usepackage{natbib}
\usepackage{amssymb}
\usepackage{wasysym}

\newcommand{\xmm}{{XMM-Newton\/}}
\newcommand{\swift}{{Swift\/}}
\newcommand{\nustar}{{NuSTAR\/}}

\newcommand{\xray}{{\hbox{X-ray}}}
\newcommand{\chandra}{{Chandra\/}}
\newcommand\iona[2]{#1$\;${\scshape{#2}}}%

\begin{document}

\title{The Curious Case of PHL 1811: Heavy Obscuration Versus Intrinsic X-ray Weakness}

\author[orcid=0000-0002-9036-0063]{B. Luo}
\affiliation{School of Astronomy and Space Science, Nanjing University, Nanjing, Jiangsu 210093, China}
\affiliation{Key Laboratory of Modern Astronomy and Astrophysics (Nanjing University), Ministry of Education, Nanjing 210093, China}
\email[show]{bluo@nju.edu.cn}  

\author{Xiaolei Chen}
\affiliation{School of Astronomy and Space Science, Nanjing University, Nanjing, Jiangsu 210093, China}
\affiliation{Key Laboratory of Modern Astronomy and Astrophysics (Nanjing University), Ministry of Education, Nanjing 210093, China}
\email{18315610282@163.com}

\author[orcid=0000-0002-9335-9455]{Jian Huang}
\affiliation{School of Astronomy and Space Science, Nanjing University, Nanjing, Jiangsu 210093, China}
\affiliation{Key Laboratory of Modern Astronomy and Astrophysics (Nanjing University), Ministry of Education, Nanjing 210093, China}
\email{jhuang@nju.edu.cn}

\author[orcid=0000-0002-0167-2453]{W. N. Brandt}
\affiliation{Department of Astronomy \& Astrophysics, 525 Davey Lab,
	The Pennsylvania State University, University Park, PA 16802, USA}
\affiliation{Institute for Gravitation and the Cosmos,
	The Pennsylvania State University, University Park, PA 16802, USA}
\affiliation{Department of Physics, 104 Davey Lab, The Pennsylvania State University, University Park, PA 16802, USA}
\email{wnbrandt@gmail.com}

\author[orcid=0000-0002-8577-2717]{Qingling Ni}
\affiliation{Max-Planck-Institut f\"{u}r extraterrestrische Physik (MPE), Gie{\ss}enbachstra{\ss}e 1, D-85748 Garching bei M\"ucnchen, Germany}
\email{qingling1001@gmail.com}

%% Use the \collaboration command to identify collaborations. This command
%% takes an optional argument that is either a number or the word "all"
%% which tells the compiler how many of the authors above the command to
%% show. For example "\collaboration[all]{(DELVE Collaboration)}" wil include
%% all the authors above this command.
%%
%% Mark off the abstract in the ``abstract'' environment. 
\begin{abstract}
{We present a systematic X-ray analysis of the narrow-line Type 1 quasar 
PHL~1811, which has long been regarded as the prototype of intrinsically
X-ray weak quasars.
A critical breakthrough came with the first detection of a 
bright X-ray flare from this source by the Einstein Probe (EP) in 2024.
We utilize archival \xray\ observations spanning 2001--2024, including the post-flare EP and Swift data.}
We confirm that PHL~1811 shows
X-ray weakness factors $f_{\rm weak} \approx 23$--$179$ across all epochs before 2024.
The 2024 EP flare marks the first detection of an X-ray nominal state
with $f_{\rm weak} \approx 0.63$, followed by a rapid flux decline.
We identify three key observational signatures that strongly support heavy obscuration:
(1) a significant hard X-ray excess above $\approx5$~keV 
in the 2015 XMM-Newton spectrum;
(2) relatively flat spectral shapes in two Swift observations; 
and (3) transitions between X-ray nominal and multiple X-ray weak states without
corresponding optical/infrared variability, consistent with 
expectations from obscuration by a clumpy dust-free absorber.
Fitting with a partial-covering obscuration model reproduces all multi-epoch spectra well.
The observed steep spectra are dominated by a small leaked/scattered fraction of 
the intrinsic continuum,
and variability is driven by changes in the leakage fraction and column density.
Our results strongly favor the scenario where PHL~1811 is obscured by
a radiatively driven accretion-disk wind from super-Eddington accretion,
unifying PHL~1811 with the broader 
population of super-Eddington accreting AGNs under a single 
obscuration framework.
\end{abstract}


%% Keywords should appear after the \end{abstract} command. 
%% The AAS Journals now uses Unified Astronomy Thesaurus (UAT) concepts:
%% https://astrothesaurus.org
%% You will be asked to selected these concepts during the submission process
%% but this old "keyword" functionality is maintained in case authors want
%% to include these concepts in their preprints.
%%
%% You can use the \uat command to link your UAT concepts back its source.
%\keywords{\uat{Galaxies}{573} --- \uat{Cosmology}{343} --- \uat{High Energy astrophysics}{739} --- \uat{Interstellar medium}{847} --- \uat{Stellar astronomy}{1583} --- \uat{Solar physics}{1476}}

%% From the front matter, we move on to the body of the paper.
%% Sections are demarcated by \section and \subsection, respectively.
%% Observe the use of the LaTeX \label
%% command after the \subsection to give a symbolic KEY to the
%% subsection for cross-referencing in a \ref command.
%% You can use LaTeX's \ref and \label commands to keep track of
%% cross-references to sections, equations, tables, and figures.
%% That way, if you change the order of any elements, LaTeX will
%% automatically renumber them.

\section{Introduction} 

X-ray observations provide one of the most powerful probes of 
accretion physics in active galactic nuclei (AGNs), systems powered 
by mass accretion onto central supermassive black holes (SMBHs). 
In the standard AGN paradigm, X-rays are produced when optical--ultraviolet (UV)
photons from the accretion disk are inverse-Compton scattered by 
relativistic electrons in a hot corona \citep[e.g.,][]{Done2010,Gilfanov2014,Fabian2017}.
A significant correlation has been observed between the
coronal X-ray emission and the accretion-disk optical/UV emission,
typically quantified by the negative
relation between the
\hbox{X-ray-to-optical} power-law slope parameter
($\alpha_{\rm OX}$)\footnote
%{$\alpha_{\rm OX}$
{It is defined as
$\alpha_{\rm OX}=-0.3838\log(f_{2500{\text{\AA}}}/f_{2\rm keV})$,
where $f_{2500{\text{\AA}}}$ and $f_{2\rm keV}$
are the \hbox{rest-frame} 2500~\AA\ and 2~keV flux densities.}
and the
2500~\AA\ monochromatic luminosity ($L_{\rm 2500{\text{\AA}}}$; e.g., \citealt{Steffen2006,Lusso2017,Huang2025}).

A small fraction of radio-quiet,
Type 1 non-broad absorption line (non-BAL) AGNs is found to be significantly
X-ray weak relative to expectations from the $\alpha_{\rm OX}$--$L_{\rm 2500{\text{\AA}}}$ relation
(e.g., $[2.7\pm0.5]\%$ among SDSS quasars
for X-ray weakness factors $f_{\rm weak}>10$; \citealt{Pu2020}).\footnote{$f_{\rm weak}$ is the factor of X-ray weakness at rest-frame 2~keV, 
relative to the expectation from the $\alpha_{\rm OX}$–$L_{2500\text{\AA}}$ relation;
a value of $f_{\rm weak}=1$ means the source follows the relation exactly.}
One main population of X-ray weak AGNs is likely related to super-Eddington accretion.
Such AGNs include weak emission-line quasars (WLQs), narrow-line Seyfert 1 
galaxies (NLS1s),
and other AGNs selected by their high accretion rates
(e.g., \citealt{Miniutti2012,Luo2015,Ni2018,Ni2020,Nardini2019,Liu2019,Liu2021,Liu2022,Komossa2020,Tripathi2020,Laurenti2022,Trefoloni2023,Yang2026}). Despite their weak X-ray emission, these objects display {typical AGN continuum
spectral energy distributions (SEDs)} from the infrared (IR) to the UV.
Some of these AGNs also exhibit {strong X-ray variability, occurring exclusively
between X-ray weak and \xray\ nominal-strength states}, with no X-ray excess states detected.
Moreover, {they do not exhibit contemporaneous optical/IR variability}, indicating a relatively
stable accretion process.
The unexpected population of ``Little Red Dots'' (LRDs),
recently discovered by the James Webb Space
Telescope (JWST),
is almost universally X-ray weak and is also considered related to super-Eddington accretion
\citep[e.g.,][]{Inayoshi2025,Maiolino2025}.

There are two competing physical interpretations for AGN X-ray weakness: heavy obscuration or intrinsic \xray\ weakness. The latter refers to a scenario where
the X-ray corona itself is intrinsically suppressed (e.g., \citealt{Leighly2007}).
X-ray spectra of X-ray weak AGNs often show signs of X-ray absorption (e.g.,
much smaller effective power-law photon indices than the canonical value of \hbox{$\sim2$}),\footnote{
{Due to generally limited photon statistics in the X-ray weak states,
spectral shapes are often characterized with
effective power-law photon indices derived
from aperture photometry.}}
and heavy X-ray obscuration
without accompanying optical/UV extinction could arise from small-scale dust-free gas near the SMBH.
We have proposed that the absorber is likely a clumpy disk wind radiatively
driven by super-Eddington accretion (e.g., \citealt{Luo2015,Ni2018,Ni2020,Liu2019,Liu2021,Wang2022}; see
Figure 1 in \citealt{Ni2018}). This model can explain, in a simple and unified manner,
the extreme X-ray weakness, extreme X-ray variability, and rather typical SEDs.
Shielding of high-energy photons by the disk wind or a thick accretion disk might
also explain the weak UV emission lines of WLQs \citep[e.g.,][]{Luo2015,Chen2024}.

On the other hand,
a key piece of evidence supporting the intrinsic X-ray weakness scenario is that
a few X-ray weak AGNs show no apparent absorption signatures during their X-ray weak states
(i.e., displaying an apparently steep power-law spectral shape with $\Gamma\sim2$).
However, Compton-thick absorption combined with a small fraction of
the intrinsic power-law continuum that is leaked or scattered is also a viable explanation for such steep spectra \citep{Wang2022}.
In this case, high-quality hard X-ray data are required to identify the $\gtrsim5$~keV excess
from the reprocessed component of the absorber (see Figure 6b in \citealt{Wang2022}).
Intrinsic \xray\ weakness also struggles to explain the variability between X-ray weak
and X-ray nominal-strength states, as
the X-ray corona would need to vary significantly, yet the accretion disk remains stable.
Additionally, no X-ray excess states have been detected in these AGNs,
which indicates that the maximum power of the corona is somehow regulated by the
$\alpha_{\rm OX}$--$L_{\rm 2500{\text{\AA}}}$ relation.
Nevertheless, intrinsic X-ray weakness remains an intriguing possibility.

PHL 1811 is an extreme quasar in many respects.
%At a redshift of
At $z=0.192$, it is unusually bright, with a $B$-band magnitude
of 13.9.
It is classified as a narrow-line Type 1 quasar with a SMBH mass of
$1.8\times10^8$~$M_{\astrosun}$
and an Eddington ratio of 1.6 \citep[e.g.,][]{Leighly2007b}.
It exhibits strong \iona{Fe}{ii} emission and weak [\iona{O}{iii}] emission in the optical spectrum 
\citep{Leighly2007b},
the Eigenvector 1 features typical of super-Eddington accretion \citep[e.g.,][]{Boroson1992,Sulentic2000,Shen2014}.
In the UV spectrum, it shows extremely weak \iona{C}{iv} emission and is classified as a WLQ.
PHL~1811 displays a typical quasar continuum SED from the IR to UV, and
it lacks strong long-term optical/IR variability (Section 3.3 of \citealt{Wang2022}).
However, it is extremely X-ray weak and variable, 
with X-ray weakness factors $f_{\rm weak}$ reaching
up to $\approx180$ (see Section~3 below), which is exceptional
even among X-ray weak AGNs. 
More importantly, in all previous \chandra\ and \xmm\ observations before 2015, its X-ray spectra were always
well described by a simple power-law model with photon indices $\Gamma\gtrsim2$ \citep[e.g.,][]{Leighly2007,Luo2015}.
Therefore, it was considered a prototypical intrinsically X-ray weak quasar \citep[][]{Leighly2007}.

Motivated by the remarkable
X-ray properties of PHL~1811,
we
investigated the X-ray properties of PHL~1811 analogs. {These analogs are 
sources sharing similar spectral features
with PHL~1811: weak UV emission lines,
large } \iona{C}{iv} blueshift, and 
strong UV \iona{Fe}{ii} and \iona{Fe}{iii} emission.
They turn out to be predominantly ($\approx94\%$) X-ray weak; however,
their X-ray weakness is likely caused by heavy absorption \citep{Luo2015,Wang2024}. These results cast doubts on
the intrinsic weakness interpretation of PHL 1811, and we speculated then that
PHL 1811
itself may also possess a heavily absorbed component 
%highly absorbed component 
that had yet to be
recognized (Section 7.2 of \citealt{Luo2015}).

A critical turning point emerged from its 2015 \xmm\ observation accompanied by a
simultaneous {NuSTAR} observation,
where
an apparent excess over a simple power-law spectrum is observed
above $\approx5$~keV.
Therefore, we proposed in \citet{Wang2022} that PHL 1811 is affected by Compton-thick absorption, and
the observed steep spectra are dominated by
a small leaked/scattered fraction of the intrinsic power-law continuum. The hard X-ray excess
from the Compton-thick absorption is only visible in the 2015 observation due to the long \xmm\ exposure
that provides the most sensitive X-ray coverage to date. 

Compared to other super-Eddington accreting AGNs whose 
X-ray weakness is well explained by the disk-wind obscuration 
scenario \citep[e.g.,][]{Liu2019,Liu2022,Ni2020,Huang2023},
a long-standing gap in the case of PHL 1811 prior to 2024 was 
the complete absence of any detection of its nominal X-ray emission, 
which would be observable when the clumpy obscuring material moves 
outside the line of sight.
This gap was finally filled on 2024 August 03 by the Einstein Probe (EP; \citealt{Yuan2022}) 
Wide-field X-ray Telescope (WXT) 0.5--4 keV
detection 
of a ``flare'' from PHL 1811 \citep{Li2024}, 
which placed the source in its expected nominal X-ray state for the first time
(see Section 3 below).
This bright state was transient and the X-ray flux dropped quickly \citep{Li2024}. 
{No subsequent X-ray flares from this source
have been reported to date
in the ongoing EP WXT all-sky survey.}
This newly emergent variability characteristic of PHL~1811 brings it 
into consistency with the behavior seen in other super-Eddington accreting AGNs.

In this paper, we present a systematic analysis of archival 
X-ray data for PHL~1811 spanning 2001--2024, 
including newly available EP and Swift observations 
obtained after the 2024 flare.
We critically examine its obscuration signatures 
and aim to unify PHL~1811 under the 
same obscuration framework as PHL 1811 analogs 
and super-Eddington accreting AGNs in general.
This paper is organized as follows.
We describe the observations and data reduction
procedures in Section 2.
In Section 3, we present results
from simple power-law spectral fitting, measurements of the 
X-ray weakness, and an analysis of the long-term X-ray variability
including the 2024 EP flare.
We present our physically motivated
obscuration modeling and related discussion
in Section~4. We summarize in Section~5.
Throughout this paper, we adopt a flat $\Lambda$CDM cosmology with
cosmological parameters of
$H_{\rm 0}=67.4~\rm km~s^{-1}~Mpc^{-1}$, $\Omega_{\rm M}=0.315$, and
$\Omega_{\Lambda}=0.685$ \citep{Planck2020}.
{All uncertainties are quoted at the 1$\sigma$ ($68.3\%$) confidence level.}

\section{X-ray OBSERVATIONS AND DATA REDUCTION}
We reduced all archival Chandra, XMM-Newton, Swift, and EP observations of PHL 1811.
The observation details are listed in Table 1. 
{The EP observations consist of a WXT survey detection of the 2024 flare,
and two subsequent observations by the Follow-up \xray\ Telescope (FXT) 
in the same year.
FXT data become publicly available after the one-year proprietary period,
while WXT data have not yet been publicly released.\footnote{
\url{https://www.cosmos.esa.int/web/einstein-probe/home}.}
We reduced the FXT data ourselves, and adopted the 0.5--4 keV flaring flux
from the WXT observation as reported in \citet{Li2024}.
The WXT observation is included in Table 1 for completeness.}
There is one additional NuSTAR observation simultaneous to the 2015 \xmm\ observation, presented
in \citet{Wang2022}. We did not reanalyze the \nustar\ data and simply refer to the results
in \citet{Wang2022}.

\begin{deluxetable*}{lrcrrrrcrrr}
\tabletypesize{\scriptsize}

\tablecaption{List of X-ray Observations and Simple Power-Law Fitting Results}
\tablehead{
\colhead{Observatory} &
\colhead{Obs. ID} &
\colhead{Obs. Date$^{a}$} &
\colhead{Exp$^{b}$} &
\colhead{Counts$^{c}$} &
\colhead{$\Gamma_{\rm PL}$$^{d}$} &
\colhead{$f_{2\rm keV}$$^{e}$}& 
\colhead{$\log L_{2500{\text{\AA}}}$$^{f}$}&
\colhead{$\alpha_{\rm OX}$}&
\colhead{$f_{\rm weak}$$^{g}$} &
\colhead{W/dof} 
%\\
%\colhead{} &
%\colhead{} &
%\colhead{} &
%\colhead{(ks)} &
%\colhead{} &
%\colhead{} &
%\colhead{($10^{-31}$ erg cm$^{-2}$ s$^{-1}$ Hz$^{-1}$)} &
%\colhead{} &
%\colhead{} &
%\colhead{}
}
\startdata
\chandra   &2957 & 2001 Dec 05   & 9.3 &79  & $1.96^{+0.21}_{-0.20}$ & 0.47& (30.93)&$-2.41^{+0.02}_{-0.03}$& $170^{+28}_{-20}$ & 58.5/55\\
\chandra   &2958 & 2001 Dec 17  & 9.8  &362  & $2.52^{+0.10}_{-0.09}$ &1.53 &(30.93) &$-2.21\pm 0.01$& $52^{+4}_{-3}$ &123.0/134 \\
\xmm       &0204310101 & 2004 Nov 01 & 18.8 &452  & $2.26\pm0.10$  & 0.48 &30.93 &$-2.41\pm 0.01$ &$166\pm 12$ & 317.0/366\\
Swift      &00030335002 & 2005 Oct 22 & 2.4 &18    & $1.80^{+0.40}_{-0.38}$ & 3.32 &30.90 &$-2.07^{+0.03}_{-0.05}$ &$23^{+9}_{-4}$ & 9.5/16\\
Swift      &00030335003 & 2006 May 12 & 1.6  &9    & $0.73^{+0.62}_{-0.69}$ & 2.58 &30.89 &$-2.11^{+0.04}_{-0.11}$& $30^{+29}_{-6}$ & 5.2/7\\
\chandra    &15357 & 2012 Nov 24  & 2.0  &72  & $2.15^{+0.21}_{-0.20}$ & 2.49 &(30.97) &$-2.15^{+0.02}_{-0.03}$&$33^{+6}_{-4}$ & 49.9/60\\
\xmm      &0761910201  & 2015 Nov 29  & 38.4   &1050 & $2.45\pm0.08$ & 0.46 &30.97&$-2.43\pm 0.01$& $179_{-7}^{+10}$&610.1/556 \\
EP WXT$^{h}$ & -- & 2024 Aug 03 & -- & -- & $2.6$ & 120 & (30.88) &$-1.47^{+0.08}_{-0.15}$ &$0.63^{+0.88}_{-0.23}$& --\\
EP FXT         &06800000032 & 2024 Aug 04   & 2.3   &137 & $2.20\pm0.19$   & 4.57 &(30.88) &$-2.01\pm 0.02$ &$17\pm2$ & 85.1/112\\
Swift      &00030335004 & 2024 Aug 06   & 2.6   &8 & $1.35^{+0.75}_{-0.72}$ & 2.50 &30.88 &$-2.11^{+0.05}_{-0.09}$&$30_{-7}^{+23}$ & 7.4/6\\
EP FXT       &06800000077  & 2024 Sep 08   & 1.4   &186 & $2.20\pm0.16$          &9.95 &(30.88) &$-1.88\pm 0.02$ &$8\pm 1$& 124.4/128\\
\enddata
\label{tbl:obslog}
\tablenotetext{a}{The observations are listed in chronological order.}
\tablenotetext{b}{Cleaned exposure time in units of ks.}
\tablenotetext{c}{Background-subtracted net counts used for spectral fitting.
{The energy ranges are 0.3--8 keV for Chandra and Swift, 0.3--10 keV for XMM-Newton,
and 0.5--8 keV for EP.}}
\tablenotetext{d}{Best-fit photon index for a simple power-law model.} 
\tablenotetext{e}{Flux density at rest-frame 2 keV in units of $10^{-31}$ erg cm$^{-2}$ s$^{-1}$ Hz$^{-1}$.
The uncertainty is not shown here, and it accounts for the $\alpha_{\rm OX}$ uncertainty.}
\tablenotetext{f}{$\log L_{2500\text{\AA}}$ in units of erg s$^{-1}$ Hz$^{-1}$. 
A number in brackets indicates that no simultaneous UV measurements were available,
and the value is adopted from a nearby \xmm\ or \swift\ observation.}
\tablenotetext{g}{Factor of X-ray weakness at rest-frame 2 keV; $f_{\rm weak}=403^{-\Delta\alpha_{\rm OX}}$.}
\tablenotetext{h}
{For the EP WXT detection of the flare, we derived $f_{2\rm keV}$ from the
0.5--4 keV flux reported by \citet{Li2024},
assuming $\Gamma=2.6$.}
\end{deluxetable*}

\subsection{Chandra Observations}
PHL~1811 was observed three times by Chandra between 2001 and 2012. 
We processed the data using the Chandra Interactive Analysis of Observations (CIAO; v4.18) 
software package. We ran the \texttt{chandra\_repro} script to generate new 
level 2 event files. We then filtered out background flares 
using the \texttt{deflare} script with an iterative 3$\sigma$ clipping algorithm.

We generated images in the 0.3--8 keV energy band using the \texttt{dmcopy} tool.
We then used the \texttt{wavdetect} tool to search for \xray\ sources 
with a false-positive probability threshold of $10^{-6}$ and wavelet scales of 
1, 1.414, 2, 2.828, 4, 5.656, and 8 pixels. 
PHL~1811 was detected in all three observations at positions 
within 0.02\arcsec--0.15\arcsec\ of its optical position. We extracted source spectra from circular regions centered on the X-ray positions with a radius of 2\arcsec. Background spectra were extracted from concentric annular regions with inner and outer radii of 6\arcsec\ and 10\arcsec, respectively. We verified that the background regions were free of contamination from other X-ray sources.

\subsection{XMM-Newton Observations}
PHL~1811 was observed twice by XMM-Newton, in 2004 and 2015. 
We used only the pn data (see Footnote 17 of \citealt{Wang2022}).
We processed the data using the XMM-Newton Science Analysis System (SAS; v22.1.0) software 
package, following the standard procedure. 
We ran the \texttt{epproc} task to produce calibrated event files.
We
generated good-time-interval files using the \texttt{tabgtigen} task
adopting a count-rate threshold of 0.4 cts s$^{-1}$, and we created cleaned 
event files using the \texttt{evselect} task.

We extracted source spectra from circular regions with a radius of 30\arcsec\ centered on the optical position of PHL~1811. Background spectra were extracted from source-free circular regions with a radius of 80\arcsec\ on the same CCD chip as the source. 

{We reduced the \xmm\ Optical Monitor (OM) 
data to obtain simultaneous
$f_{2500{\text{\AA}}}$ measurements. 
Five OM filters were used in the 2015 observation, including UVW2, UVM2, UVW1, $U$, and $B$, with effective wavelengths of $2120\,$\AA, $2320\,$\AA, $2910\,$\AA, $3440\,$\AA, and $4500\,$\AA, respectively.
Only the UVM2 filter was used in the 2004 observation. 
We utilized the pipeline \textsc{omichain} to process the OM data for each \hbox{XMM-Newton} observation.
For each filter in every observation, we extracted the mean magnitude and flux density of all exposures from the merged source list (OBSMER file) produced by the pipeline.
We corrected for Galactic extinction
using the dereddening law of \citet{Cardelli1989}
and \citet{ODonnell1994} with $E(B-V) = 0.048$ \citep{Schlegel1998}.
For the 2015 observation, we derived $f_{2500{\text{\AA}}}$
by interpolating between the UVW1 and $U$-band flux densities.
For the 2004 observation, we adopted the measurement
% (updated to our cosmological parameters)
from \citet{Leighly2007b},
which was derived from the 2001 HST spectrum
that shows good consistency with the OM UVM2 measurement.}

\subsection{Swift Observations}
PHL~1811 was observed three times by the Swift X-ray Telescope (XRT) between 2005 and 2024; 
the 2024 observation is a follow-up of the EP flare \citep{Li2024}. All observations were performed in Photon Counting (PC) mode. We processed the data using the \texttt{xrtpipeline} task 
in the High Energy Astrophysics Software (HEASoft; v6.33) package.

We searched for X-ray detections of PHL~1811 in 
the 0.3--8 keV images using \texttt{wavdetect} with the same setup as for the 
Chandra observations. It was detected in the two earlier observations (2.0\arcsec\ and 
4.5\arcsec\ offsets from the optical position) but not in
the 2024 observation; \citet{Li2024} also reported a nondetection for the 2024 observation
and provided a flux upper limit. However, it was detected in the 2024 observation 
(positional offset of 3.6\arcsec) if 
we relax 
the false-positive probability threshold of \texttt{wavdetect} from $10^{-6}$
to $10^{-5}$. 
{This $10^{-5}$ threshold is entirely
appropriate for targeted source
searching at a pre-specified
position, rather than blind
field-wide source searching.}
We then performed aperture photometry (with the same source and
background regions used for spectral extraction below) to assess 
the significance of this detection, adopting the binomial no-source probability ($P_{\rm B}$)
parameter
(e.g., Equation 1 of \citealt{Luo2015}). The resulting $P_{\rm B}$ value is 
$1.1 \times 10^{-5}$, corresponding to a $\approx4.4\sigma$ detection. Thus we consider
PHL~1811 also detected in the 2024 Swift observation.

We extracted source spectra from circular regions with a radius of 
20\arcsec\ centered on the optical position. Background spectra 
were extracted from concentric annular regions with inner and 
outer radii of 40\arcsec\ and 90\arcsec. We generated the auxiliary response files (ARFs) using the \texttt{xrtmkarf} task and obtained the response matrix files (RMFs) from 
the 
calibration database (CALDB).

{We reduced the \swift\ UVOT data to obtain simultaneous
$f_{2500{\text{\AA}}}$ measurements.
The UVOT has six filters ($V$, $B$, $U$, UVW1, UVM2, and UVW2),
with slightly different effective
wavelengths (5402\,\AA, 4329\,\AA, 3501\,\AA, 2634\,\AA, 2231\,\AA, and 2030\,\AA)
compared to the \xmm\ OM. For the 2005 observation, only the UVW2, UVM2, and
UVW1 filters were used, whereas the 2006 and 2024 observations employed all
six filters. After aspect correction, data from each segment in each
filter were \hbox{co-added} using the task \textsc{uvotimsum}. The task
\textsc{uvotdetect} was then run for each image to detect sources.
Source counts were extracted from a 5\arcsec\ circular region
centered on the source position determined by \textsc{uvotdetect}.
Background counts were extracted from a nearby \hbox{source-free}
region of radius 15\arcsec. Finally, the magnitudes and fluxes
were calculated using the task \textsc{uvotsource}. The results were
then corrected for the Galactic extinction. Using these UVOT photometric measurements,
we derived $f_{2500{\text{\AA}}}$ values by interpolation or extrapolation as needed.}

\subsection{EP Observations}
\begin{figure}
\centering
\includegraphics[width=1\linewidth]{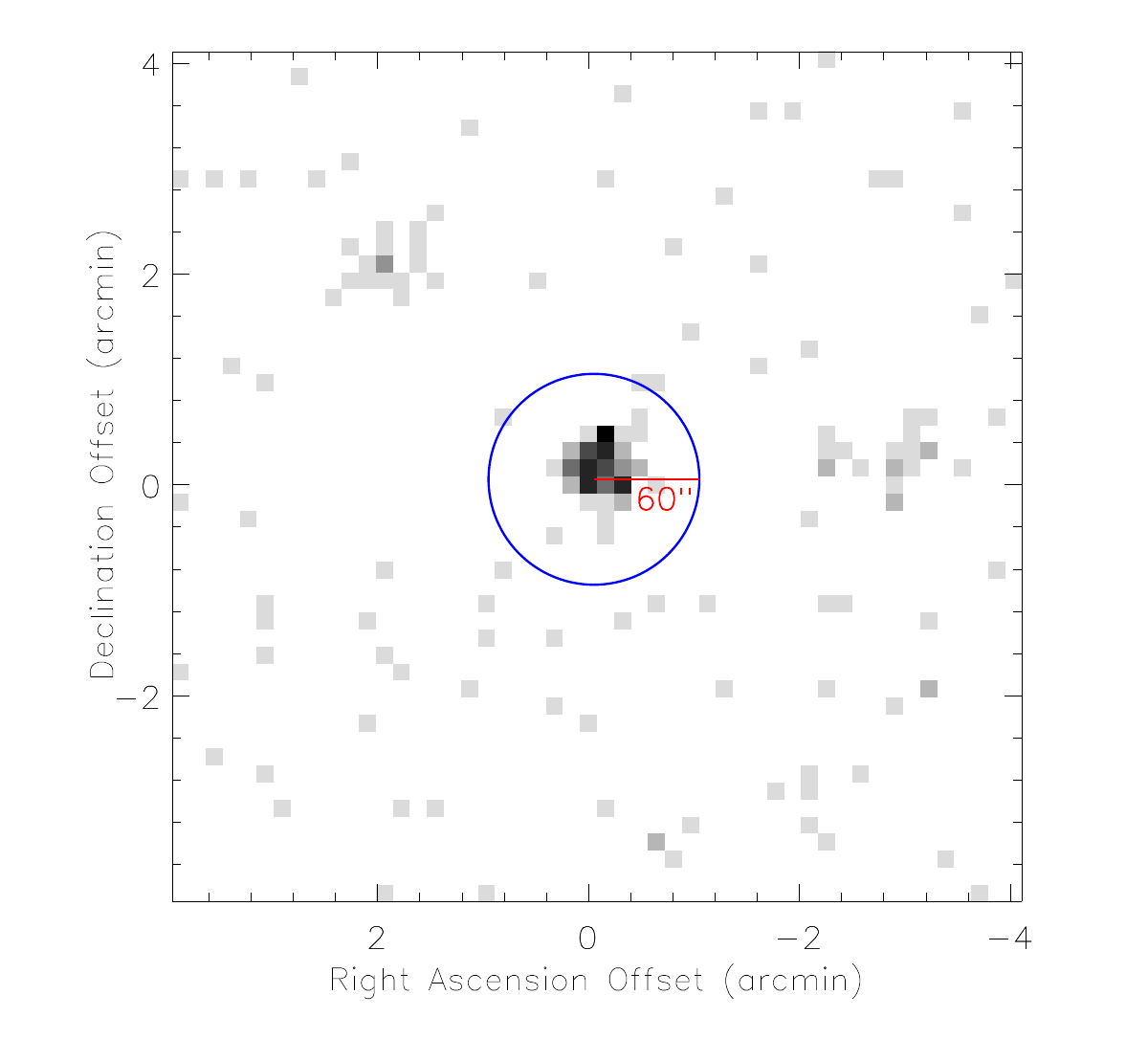}
\caption{EP FXTA image in the 0.5--8 keV band for the 2024 August
observation. PHL 1811 is clearly detected. The blue circle marks the source
extraction region, centered on the optical position. 
The background extraction region is an annulus with inner and outer radii of 100\arcsec\ 
and 160\arcsec,
respectively. Two additional X-ray sources (to the upper left and right of PHL~1811) 
partially overlap
the background region and were excluded during background extraction.}
\label{ximg}
\end{figure}

{As noted previously, we only reduced the publicly available EP FXT observations.}
PHL~1811 was observed twice by the EP FXT in 2024, 
on August 4 and September 8, and the data were retrieved from the 
EP public archive.\footnote{\url{https://ep.bao.ac.cn/ep/next/data-release}.} 
We processed the data using the FXT 
Data Analysis Software (FXTDAS; v1.20) package. 
We ran the \texttt{fxtchain} task in Full Frame (FF) mode to generate calibrated event files.
For each of the FXT modules (FXTA and FXTB) in each observation,
we searched for an X-ray detection of PHL~1811 in
the \hbox{0.5--8 keV} image using \texttt{wavdetect} with the same setup as for the
Chandra observations. PHL~1811 was detected in all images with
positional offsets of $\approx1.5\arcsec$--$4.0\arcsec$. The \hbox{0.5--8~keV} 
FXTA image from the 2024 August observation is shown in 
Figure~\ref{ximg}. {PHL~1811 has $\approx 69$ net counts (after subtracting $\approx 5$ background
counts) within a 60\arcsec-radius aperture in this image.} 

We extracted source spectra from circular regions with a radius of 
60\arcsec\ centered on the optical position.
Background spectra were extracted from concentric 
annular regions with inner and outer radii of 100\arcsec\ and 160\arcsec, respectively. 
Two additional X-ray sources partially overlap
the background regions; we excluded them from the background extraction
using circular apertures with a radius of 40\arcsec.
For each observation, we combined the FXTA and FXTB spectra using the \texttt{addspec} 
tool in HEASoft.

\section{SIMPLE POWER-LAW SPECTRAL ANALYSIS and X-ray properties}
We used XSPEC (v12.14.0; \citealp{Arnaud1996}) to fit the 0.3--8 keV 
Chandra, 0.3--10 keV \xmm, 0.3--8~keV Swift, and 0.5--8 keV EP spectra.
All \xray\ spectra were grouped to have at least one count per bin, and we used the W 
statistic\footnote{\url{
https://heasarc.gsfc.nasa.gov/docs/xanadu/xspec/manual/XSappendixStatistics.html}.}
in XSPEC.
We adopted a simple power-law model to characterize the 
basic X-ray properties of PHL~1811 without invoking complex physical assumptions. 
The XSPEC model is 
\texttt{phabs*zpowerlaw}, where the Galactic neutral hydrogen column density was fixed to $N_{\rm H, Gal} = 4.22 \times 10^{20}\ \text{cm}^{-2}$ \citep{HI4PI2016}.

The best-fit results are summarized in Table 1. 
The \chandra\ and \xmm\ best-fit $\Gamma$ values are 
consistent with previous measurements for the same observations within the uncertainties
\citep{Leighly2007,Wang2022}.
The simple 
power-law model describes the spectra adequately (reduced W statistic $\lesssim1$) 
except for the 2015 \xmm\ spectrum (reduced W statistic $610.1/556$).
The 2015 \xmm\ spectrum exhibits significant 
residuals above $\approx5$~keV \ (Figure 4 of \citealt{Wang2022}), but these residuals
do not significantly affect the $f_{2\rm keV}$ or
{$\alpha_{\rm OX}$} measurements.
{The best-fit $\Gamma$ values are generally large ($\gtrsim2$), with a median 
value of 2.2 for all the observations. However, the 
2006 and 2024 \swift\ observations imply flatter spectral shapes,
with $\Gamma=0.73^{+0.62}_{-0.69}$ and $1.35^{+0.75}_{-0.72}$, respectively.
For the 2006 observation, the $90\%$ and $95.4\%$ (2$\sigma$) confidence level
upper bounds on $\Gamma$ are 1.73 and 1.94, respectively.
The best-fit $\Gamma$ thus differs from the median $\Gamma$ at the $\gtrsim2\sigma$ level.}
We note that the 3--24 keV effective photon index\footnote{The \nustar\ effective photon index ($\Gamma_{\rm eff}$)
was determined from the ratio between the 8–24~keV and 3–8~keV counts,
assuming a simple power-law spectrum.} 
derived from the 2015 \nustar\
aperture photometry ($\Gamma_{\rm eff}=1.4^{+0.8}_{-0.7}$) is also relatively flat
\citep[Section 3.2 of][]{Wang2022}. 

To quantify the X-ray weakness of PHL~1811, we first computed 
the $\alpha_{\text{OX}}$ parameter (Table~1).
{We used the simultaneous $f_{2500{\text{\AA}}}$ measurements for the 
\xmm\ and \swift\ observations (Section~2). For the \chandra\ and EP
observations lacking
simultaneous UV data, we adopted values from nearby
\xmm\ or \swift\ measurements. The
$\log L_{2500{\text{\AA}}}$ values are listed in Table~1, differing by at most
$\approx23\%$.}
We then utilized the 
\citet{Pu2020} 
$\alpha_{\rm OX}$--$L_{\rm 2500{\text{\AA}}}$ 
relation\footnote{The $\alpha_{\rm OX}$--$L_{\rm 2500{\text{\AA}}}$ relation 
exhibits an apparent luminosity dependence,
with the slope of
the relation being steeper for more luminous quasars 
\citep[e.g.,][]{Steffen2006,Huang2025}.
The \citet{Pu2020} relation was derived from a 
quasar sample including a substantial fraction of 
sources with luminosities comparable 
to that of PHL~1811.
Adopting a different relation would yield
slightly different X-ray weakness factor measurements,
but these differences do not materially affect our subsequent discussion.}
to compute the deviations ($\Delta\alpha_{\rm OX}$)
between the
observed $\alpha_{\rm OX}$ values 
and the expectation from this relation.
The factor of \xray\ weakness at rest-frame 2~keV, relative to the expectation from the $\alpha_{\rm OX}$–$L_{2500\text{\AA}}$ relation, 
is then given by $f_{\rm {weak}} = 403^{-\Delta\alpha_{\text{OX}}}$.
The corresponding $f_{\rm weak}$ values are listed in Table~1.

{In addition, for the EP WXT flare detection,
only an unabsorbed 0.5--4 keV flux was reported by \citet{Li2024}.
We thus derived $f_{2\rm keV}$ from the 0.5--4 keV flux assuming $\Gamma=2.6$,
taken as the intrinsic power-law photon index (see \citealt{Wang2022} and
Section 4 below). The resulting $\alpha_{\text{OX}}$
is $-1.47^{+0.08}_{-0.15}$, corresponding to
$f_{\rm weak}=0.63^{+0.88}_{-0.23}$. We note that adopting an alternative
$\Gamma$ does not significantly change our results given the large
measurement uncertainties. For example, $\Gamma=1.8$ yields $\alpha_{\text{OX}}=-1.44^{+0.08}_{-0.15}$ and $f_{\rm weak}=0.52^{+0.73}_{-0.19}$.
The typical scatter of the
$\alpha_{\rm OX}$–$L_{2500{\text{\AA}}}$ relation is $\approx0.15$
\citep[e.g.,][]{Steffen2006,Huang2025}, corresponding to $f_{\rm weak}$ between 0.41 and 
2.46.}
Therefore, the X-ray emission strength during
this flaring state is consistent with the expectation, and PHL 1811 was observed
for the first and only time in its X-ray nominal state.

\begin{figure*}
\centering
\includegraphics[trim=0 0 0 0,clip, width=0.48\linewidth]{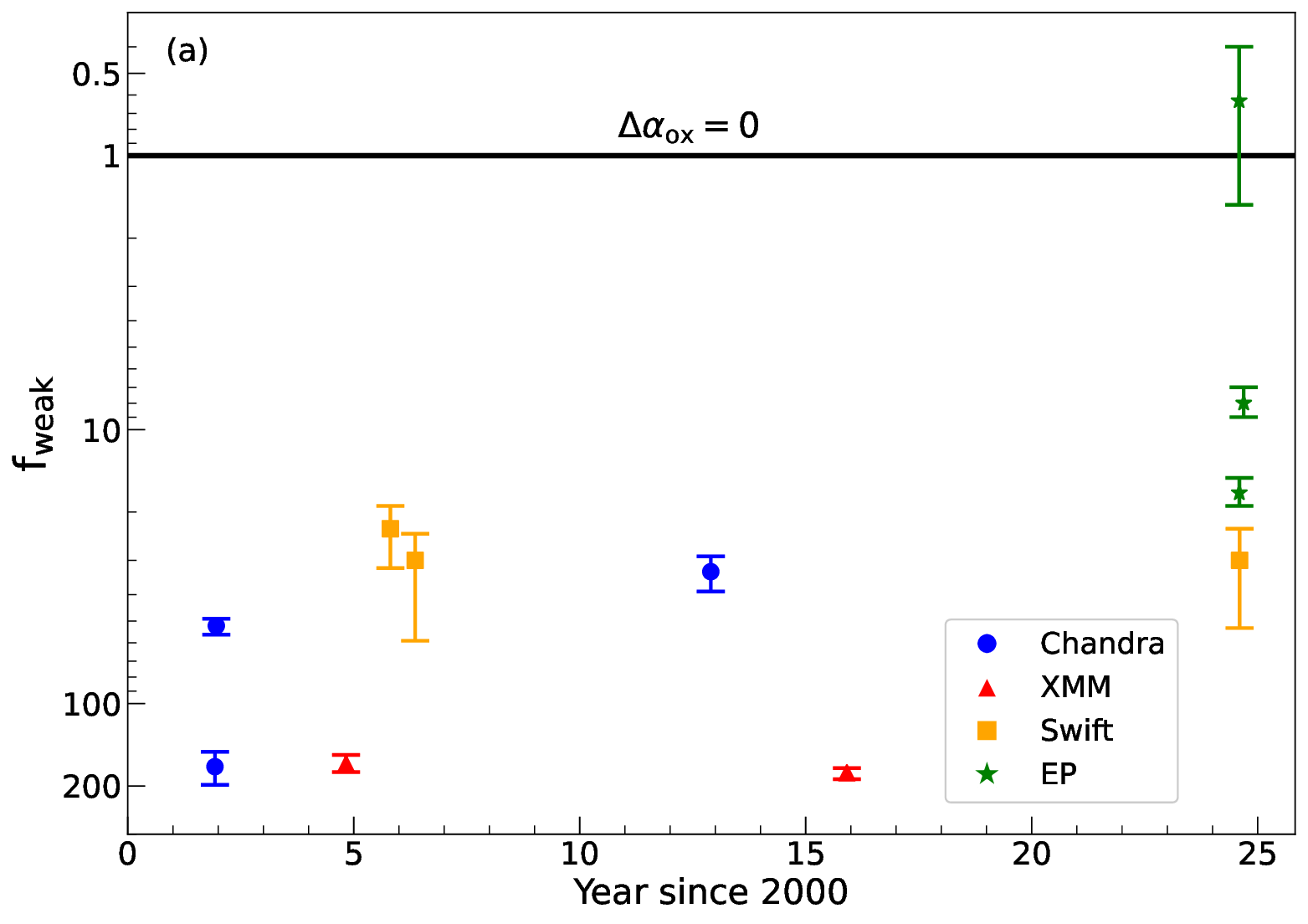}
\includegraphics[trim=0 0 0 0,clip, width=0.48\linewidth]{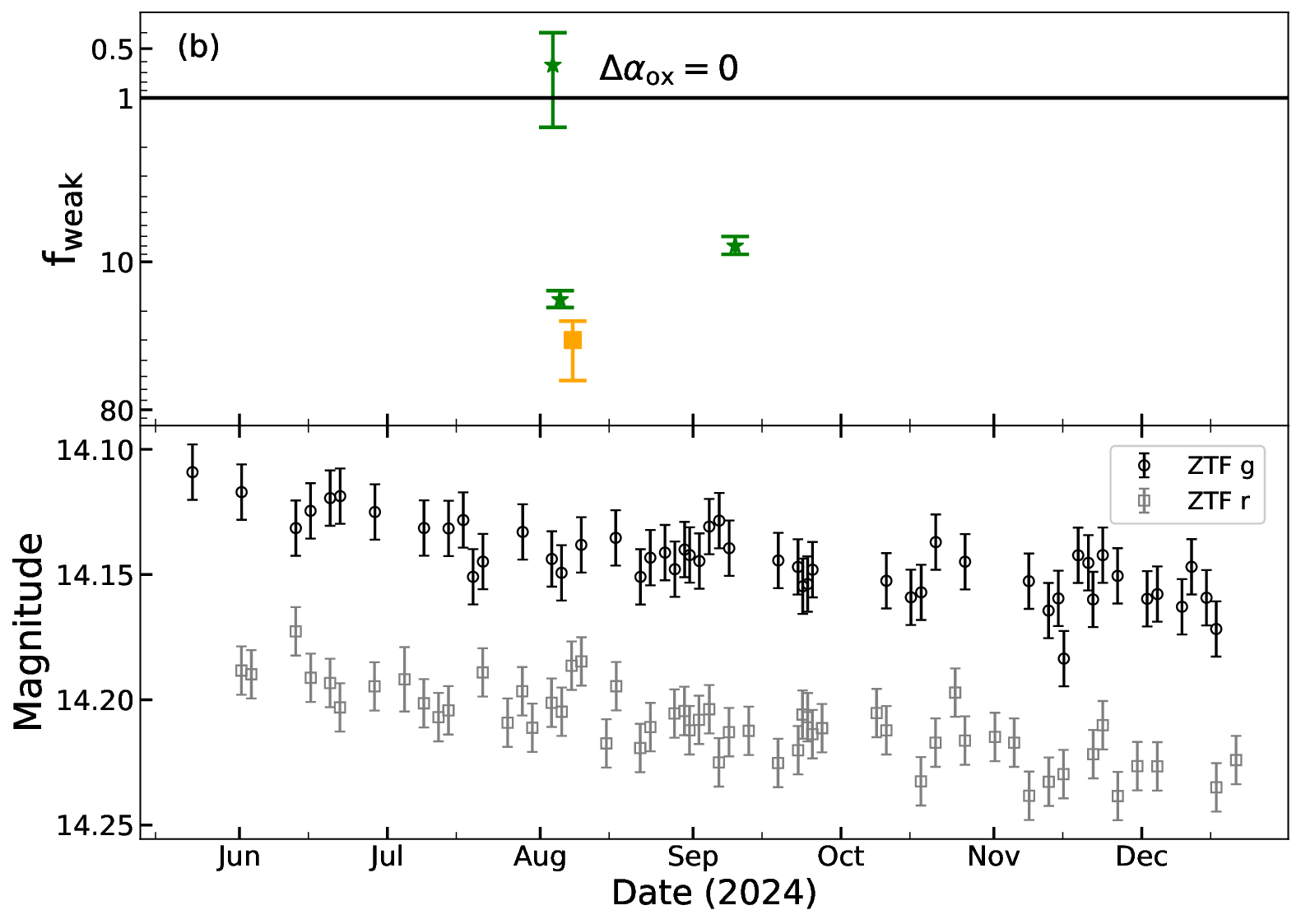}

\caption{(a) X-ray weakness factors ($f_{\rm weak}$) of PHL~1811 at different epochs. The 
Chandra, XMM-Newton, Swift, and EP FXT $f_{\rm weak}$ values 
were derived from spectral fitting with a simple power-law model 
(Table~\ref{tbl:obslog}). The EP WXT $f_{\rm weak}$ value, corresponding to the flaring state,
was derived from the flux measurement reported in \citet{Li2024}. 
The solid line indicates $f_{\rm weak}$ =1 ($\Delta\alpha_{\rm OX} = 0$), representing the X-ray 
nominal-strength state. 
PHL~1811 was extremely X-ray weak and variable, 
and it was observed
for the first and only time in its X-ray nominal state during the 2024 EP flare.
The flux declined rapidly after the flare, and it recovered somewhat about
one month later. (b) Top: Zoomed-in view of {the temporal evolution
of $f_{\rm weak}$} during the 2024 EP flaring epoch,
including the flare and its follow-up observations. Bottom: ZTF $g$- and $r$-band light
curves over the same epoch, with intraday measurements binned. In contrast to the extreme X-ray 
variability, the maximum optical variability amplitude is
only $\approx0.07$ mag ($\approx7\%$ relative flux change).}
\label{lc}
\end{figure*}

{We present the temporal evolution of the \xray\ weakness factor ($f_{\rm weak}$)
for PHL~1811 in Figure~\ref{lc}.}
Prior to 2024, PHL~1811 exhibited extreme X-ray weakness and variability, with 
$f_{\rm weak}$ ranging from $\approx23$ to $\approx179$. It also displayed rapid
variability between the two \chandra\ observations in 2001, varying by a factor
of $\approx3.3$ in 12 days. This property was used to argue against a 
heavily obscured
spectrum dominated by distant reflection/scattering \citep{Leighly2007}.
The 2~keV flux density during the 2024 EP flare was $\approx284$ times higher
than the previous 2015 XMM-Newton measurement, and $f_{\rm weak}$ 
plummeted from
$\approx179$ to $\approx0.63$. PHL~1811 was in its X-ray nominal state
during the flare. 
This state appears transient: the
2~keV flux density dropped by a factor of $\approx27$ ($f_{\rm weak}\approx17$)
in the EP FXT follow-up observation
33~hours later, and by a factor of $\approx48$ ($f_{\rm weak}\approx30$)
in the \swift\ follow-up observation
87 hours later. The flux partially recovered ($f_{\rm weak}\approx8$) in the subsequent EP FXT observation approximately one month later.

{Extreme X-ray variability without contemporaneous significant optical/IR variability is a 
characteristic feature of super-Eddington accreting AGNs that show extreme X-ray variability 
\citep[e.g.,][]{Liu2019,Liu2022,Ni2020,Huang2023}. 
We thus constructed and examined optical and IR light curves using
archival data from the Zwicky Transient Facility (ZTF;
\citealt{Masci2019}) 
in the $g$ and $r$ bands, as well as the Near-Earth Object Wide-field Infrared Survey Explorer
Reactivation Mission
(NEOWISE; \citealt{Mainzer2014}) in the W1 (3.4\,$\mu$m) and W2 (4.6\,$\mu$m) bands.
{The ZTF measurements were corrected for the Galactic extinction.}
The maximum ZTF $g$- or $r$-band variability amplitude is $\approx0.21$ mag ($\approx21\%$ 
relative flux change) 
over the long-term baseline between 2018 and 2025.
The ZTF light curves fully cover the 2024 EP flaring epoch (see Figure~\ref{lc}b), 
and the maximum variability amplitude is 
$\approx0.07$ mag ($\approx7\%$) between 2024 May 22 and 2024 December 20.
For the NEOWISE light curves, the maximum variability amplitude is
$\approx0.16$ mag ($\approx16\%$) between 2014 and 2024.
NEOWISE ceased operations before the 2024 EP flare.
We therefore confirm that PHL~1811 did not exhibit significant optical variability
contemporaneous with the 2024 EP flare.}

\section{DISCUSSION}

%Unifying PHL~1811 and super-Eddington accreting AGNs under the
%obscuration scenario}

Motivated by the hard X-ray excess detected in the 2015 
\xmm\ and \nustar\ observations \citep{Wang2022} and 
the bright 2024 EP flare event, we systematically analyzed all available 
archival \chandra, \xmm, \swift, and EP observations of PHL~1811. 
We confirmed its extreme X-ray weakness and strong variability 
across all epochs before 2024, and we found that the 2024 EP flare marks the first 
and only time that PHL~1811 has been observed in its X-ray nominal state. 
Combined with its X-ray and multiwavelength properties reported in the literature, we summarize below the observational signatures of obscuration in PHL~1811.

\begin{enumerate}
\item 
Its 2015 \xmm\ and \nustar\ spectra exhibit significant residuals
above $\approx5$~keV relative to a simple power-law model, which cannot be
explained by an intrinsically weak power-law continuum plus 
a typical Compton-reflection component.
Instead, partial-covering Compton-thick obscuration
can naturally explain these spectra: soft X-rays are dominated by a leaked component,
while the hard X-rays are dominated by a reprocessed component
from the obscuring material (Section~4.1 of \citealt{Wang2022}).

\item 
The 2015 \nustar\ 3--24~keV effective photon index
($\Gamma_{\rm eff}=1.4^{+0.8}_{-0.7}$), as well as the 2006 and 2024 \swift\
0.3--8~keV best-fit power-law photon indices
($\Gamma=0.73^{+0.62}_{-0.69}$ and $1.35^{+0.75}_{-0.72}$),
are flatter than the median $\Gamma$ value of $2.2$ for all the observations.
While we cannot rule out a scenario where the intrinsic continuum 
is variable in both flux and spectral shape, flat spectral shapes are 
naturally expected in the obscuration scenario. Under 
a partial-covering heavy obscuration scenario, the 
$\approx0.3$--8~keV spectral shape may remain steep in weak 
states if $N_{\rm H}$ is large (e.g., Compton-thick) 
and the leaked component dominates 
(c.f. Figure~6b of \citealt{Wang2022}); it becomes flatter if $N_{\rm H}$ is relatively small and the transmitted (absorbed) component emerges.

\item
The variability between an X-ray nominal state and multiple X-ray weak states
is consistent with expectations from obscuration by a clumpy absorber.
The nominal state appears when the absorber moves out of the line of sight,
or when the X-ray corona is observed through a ``hole'' in the absorber.
Rapid variability is explained by variations 
in the covering fraction (leaked fraction)
of a small-scale clumpy absorber. Moreover, X-ray variability 
without contemporaneous optical/IR variability can be naturally
explained by
a small-scale, dust-free absorber.

\end{enumerate}

More importantly, except for its more extreme \xray\ weakness,
PHL~1811 now exhibits X-ray and multiwavelength characteristics
similar to those of WLQs, PHL~1811 analogs, and other super-Eddington accreting AGNs.
It is thus natural to consider unifying these objects
under the same obscuration framework. 
We note that it is not feasible to
rigorously rule out a scenario where there is both intrinsic X-ray weakness
and obscuration, as there is generally always flexibility to accommodate a weaker
intrinsic continuum given the current spectral quality. However,
Occam’s razor would favor a simpler obscuration scenario without invoking
intrinsic X-ray weakness, and this scenario appears sufficient
to explain the observed X-ray and multi-wavelength properties in general.

We then analyzed the multi-epoch X-ray spectra of PHL~1811 using a 
physically motivated partial-covering 
obscuration model, following the approach described in 
Section 4.2 of \citet{Wang2022}.
The observed X-ray spectrum is the superposition of three components:
\begin{enumerate}
    \item 
A small fraction ($f_{\rm leak}$) of the 
intrinsic continuum that leaks through the 
clumpy absorber or is scattered by distant ionized gas.

\item A transmitted component 
that passes through the obscuring gas 
with column density $N_{\rm H}$, 
undergoing photoelectric absorption and Compton scattering.

\item A reprocessed component produced 
when intrinsic X-rays are Compton-scattered by the obscuring material.

\end{enumerate}
The corresponding XSPEC model is:
\[
\texttt{phabs} \times (\texttt{c1} \times \texttt{zpow} + \texttt{c2} \times \texttt{cabs} \times \texttt{zphabs} \times \texttt{zpow} +\texttt{borus02})
\]
where \texttt{c1} 
represents the leakage fraction $f_{{\rm leak}}$, $\texttt{c2} = 1 - \texttt{c1}$,
and \texttt{borus02} is the model from \citet{Balokovic2018} used here to 
simulate reprocessed emission from the clumpy obscuring material.
As discussed in Section 4.2 of \citet{Wang2022}, this 
borus02-based partial-covering obscuration model 
has several inherent caveats. Although it cannot fully 
reproduce the detailed physics of clumpy disk wind obscuration, 
it works reasonably well for 
fitting the multi-epoch spectra of PHL~1811 presented here.

A key assumption of the obscuration scenario is 
that the intrinsic coronal X-ray emission 
is stable, with variability amplitudes not exceeding the $\approx20\%$--$50\%$ 
range typical of radio-quiet Type 1 
AGNs (e.g., \citealt{McHardy2006,MacLeod2010,Yang2016,Timlin2020}). 
Accordingly, we fixed the normalization of the 
intrinsic power-law continuum to the value 
measured by EP WXT during the nominal X-ray state of PHL~1811.
Considering the large uncertainties of the EP WXT measurement, 
we also verified that varying this normalization parameter by a factor of 2 
does not significantly affect our derived obscuration parameters.

The main parameters we seek are the leakage fraction ($f_{\rm leak}$) 
and the column density ($N_{\rm H}$) of the absorber, 
whose variations are likely responsible for the 
observed extreme X-ray variability. 
The other two free parameters are the intrinsic 
power-law photon index ($\Gamma$) and the opening 
angle of the absorber ($\cos\theta_{\rm oa}$) in \texttt{borus02}.
We first jointly fitted the 2001 December 17 \chandra\ spectrum 
and the two \xmm\ spectra, 
which have $\approx360$--$1050$ net spectral counts, 
to determine $\Gamma$ and $\cos\theta_{\rm oa}$ 
(along with the $f_{\rm leak}$ and $N_{\rm H}$ 
values for these three observations). 
We then fixed these two parameters at their best-fit 
values and performed individual fits for 
the remaining lower-quality spectra, 
obtaining the $f_{\rm leak}$ and $N_{\rm H}$ values for all epochs.

\begin{deluxetable*}{lcrcrcrrr}

\tablecaption{Best-fit Parameters for the Partial-Covering
Obscuration Model}
\tablehead{
\colhead{Observatory$^{a}$} &
\colhead{Obs. Date} &
\colhead{$f_{\rm weak}$$^{b}$} &
\colhead{$\cos\theta_{\rm oa}$$^{c}$} &
\colhead{$f_{\rm leak}$} &
\colhead{$\Gamma$$^{c}$} &
\colhead{$\log N_{\rm H}$ (cm$^{-2}$)$^{d}$} &
\colhead{Main~Com.$^{e}$} &
\colhead{W/dof} 
%\colhead{} &
%\colhead{} &
%\colhead{} &
%\colhead{} &
%\colhead{} &
%\colhead{} &
%\colhead{(cm$^{-2}$)} &
%\colhead{}&
%\colhead{}
}
\startdata
\chandra\ & 2001 Dec 17 &52  & $0.58\pm0.01$ & $(1.2\pm0.1)\times10^{-2}$ & $2.6_{-0.05}$ & $25.5_{-0.6}$ & leak, rep & 121.4/134\\
\xmm\     & 2004 Nov 01 &166  & -- & $(3.0\pm0.2)\times10^{-3}$ & -- & $25.5_{-0.2}$ & leak, rep & 327.8/368\\
\xmm\     & 2015 Nov 29 &179 & -- & $(3.3\pm0.1)\times10^{-3}$ & -- & $25.5_{-0.2}$ & leak, rep & 556.3/556\\
\hline
\chandra\ & 2001 Dec 05 &170 & $0.58$ & $(2.4\pm0.3)\times10^{-3}$ & 2.6 & $24.4^{+0.7}_{-0.2}$ & leak, rep & 60.3/53 \\
\chandra\ & 2012 Nov 24 &33 & -- & $(1.7\pm0.2)\times10^{-2}$ & -- & $25.1_{-0.7}$ & leak, rep &54.3/60\\
Swift     & 2005 Oct 22 &23  & -- & $2.1^{+0.6}_{-0.5}\times10^{-2}$ & -- & $24.4_{-0.4}$ & leak, rep &12.6/16\\
Swift     & 2006 May 12 &30 & -- & $9.2^{+4.8}_{-3.7}\times10^{-3}$ & -- & $23.6\pm0.2$ & leak, tra &7.6/7\\
Swift     & 2024 Aug 06 &30  & -- & $1.1^{+0.5}_{-0.4}\times10^{-2}$ & -- & $23.8\pm0.2$ & leak, tra &6.0/6\\
EP        & 2024 Aug 04 &17  & -- & $(3.1\pm0.3)\times10^{-2}$ & -- & $24.2^{+0.5}_{-0.2}$ & leak, rep &85.4/112\\
EP        & 2024 Sep 08 &8  & -- & $(6.9\pm0.5)\times10^{-2}$ & -- & $24.5_{-0.6}$ & leak &130.1/128\\
\enddata
\label{tbl:obsmod}
\tablenotetext{a}{Results for the three jointly fitted spectra are listed first, followed by remaining observations ordered by observatory: Chandra, Swift, and EP.}
\tablenotetext{b}{X-ray weakness factors, identical to those listed in Table~1 for quick reference.}
\tablenotetext{c}{Opening angle and intrinsic photon index are free but tied in the joint fit for the first three entries, and fixed at the joint-fit best-fit values for all others. The best-fit $\Gamma$ has no upper uncertainty, as 2.6 is the upper limit of the \texttt{borus02} model.}
\tablenotetext{d}{$N_{\rm H}$ values without 
upper uncertainties indicate that 
the corresponding spectra provide no constraint on higher column densities.}
\tablenotetext{e}{The dominant component(s) in the observed X-ray 
spectrum: ``leak'' refers to the continuum leaked through gaps in the 
clumpy absorber or scattered by distant ionized gas; 
``tra'' refers to the transmitted component absorbed 
by the obscuring material; 
``rep'' refers to the Compton-reprocessed component from the obscuring material.}
\end{deluxetable*}

\begin{figure}
\centering
\includegraphics[trim=0 0 0 0,clip, width=1.1\linewidth]{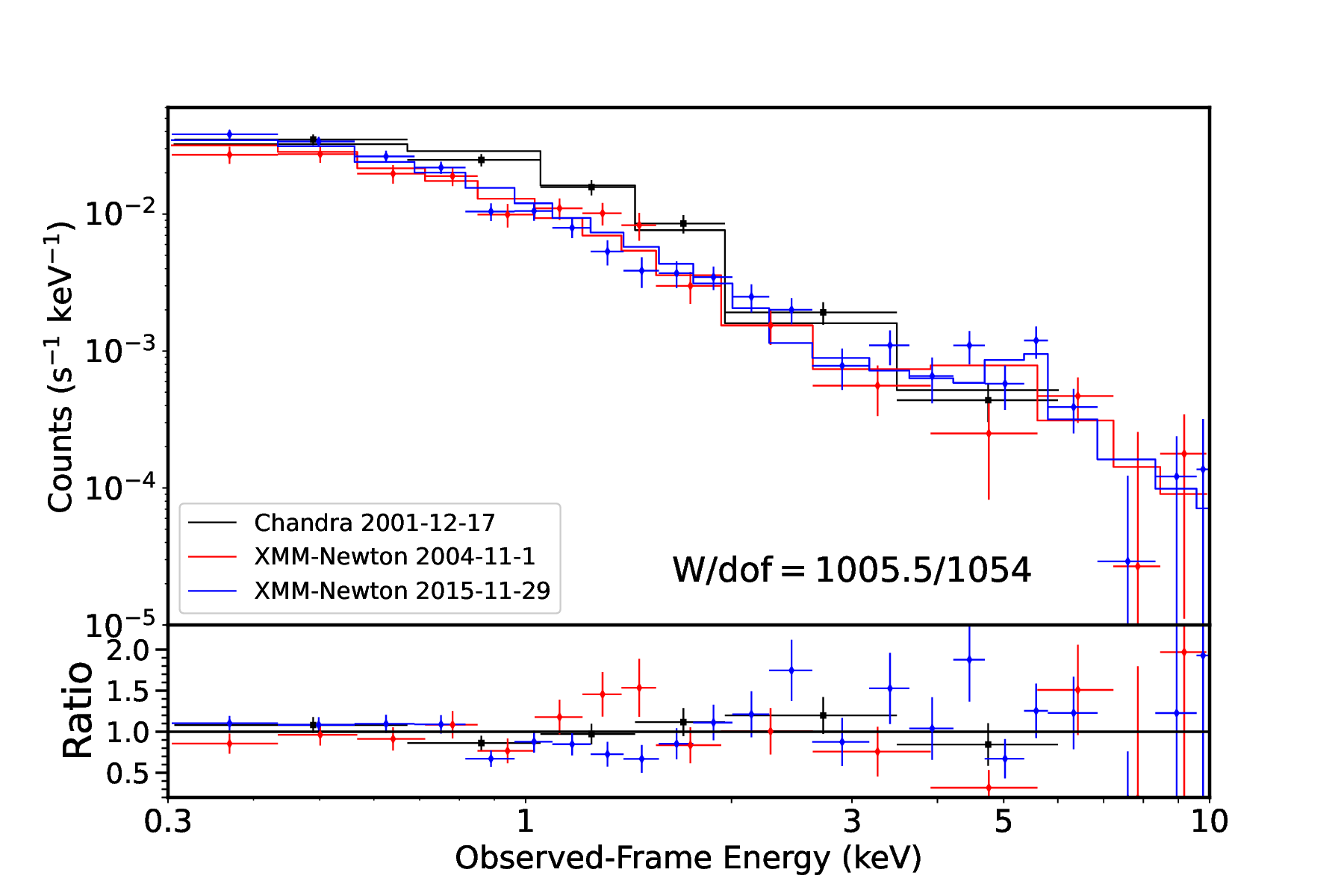}

\caption{The three jointly fitted \chandra\ and \xmm\ spectra of PHL~1811, overlaid with the best-fit partial-covering obscuration model. The bottom panel displays the data-to-model ratios for each spectrum.
The spectra are grouped for display purposes only.}
\label{joint}
\end{figure}

The best-fit results are presented in Table~\ref{tbl:obsmod}. 
We first listed the results for the three jointly fitted spectra, 
and then the remaining observations grouped by observatory
in the order of Chandra, Swift, and EP. The three jointly fitted spectra are shown
in Figure~\ref{joint}. 
Overall, this partial-covering obscuration model 
provides a good description of all multi-epoch spectra, 
as indicated by the fitting statistics in 
Table~\ref{tbl:obsmod} and the residuals in 
Figure~\ref{joint}. 
In particular, the fit to the 2015 \xmm\ spectrum 
is significantly improved compared to 
the simple power-law model, with 
the reduced W statistic decreasing from $610.1/556$ to $556.3/556$.

\begin{figure*}
\centering
\includegraphics[trim=0 0 0 0,clip, width=1\linewidth]{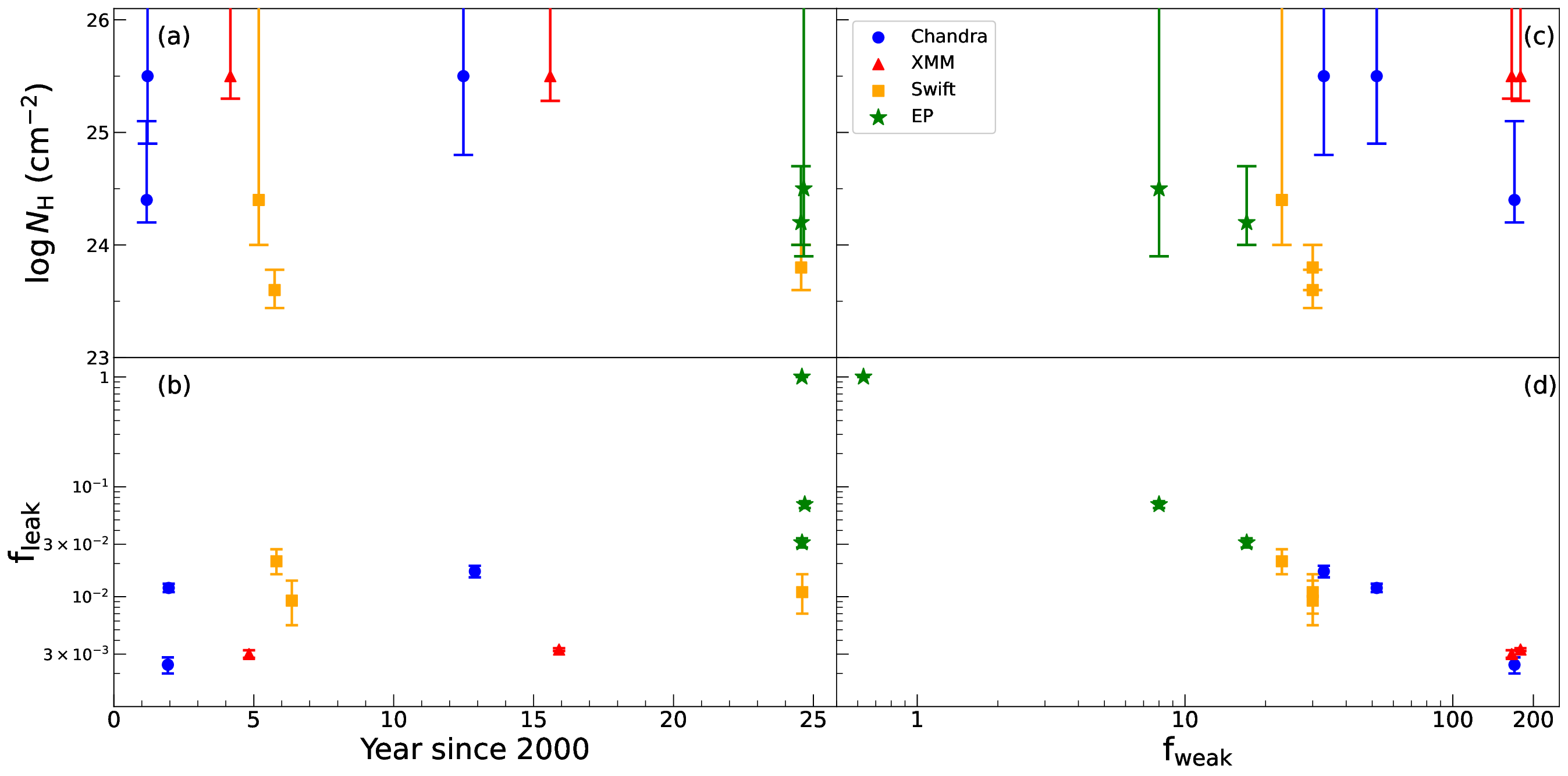}

\caption{Long-term evolution of the (a) column 
density ($\log N_{\rm H}$) and (b)
leakage fraction ($f_{\rm leak}$) of the absorber in PHL~1811, 
alongside $f_{\rm weak}$
plotted against the (c)
column density and (d)
leakage fraction.
The EP WXT flare is not shown in the $\log N_{\rm H}$ panels, and we set its $f_{\rm leak}$ to unity, corresponding to the unobscured X-ray nominal state. 
There appears to be a significant correlation between $f_{\rm leak}$ and $f_{\rm weak}$,
indicating that the observed weak-state spectra are largely dominated by the 
leaked component; the 2006 and 2024 Swift observations have the lowest $N_{\rm H}$ values,
and they deviate slightly from the $f_{\rm leak}$ vs.\ $f_{\rm weak}$ relation.}
\label{leakandnh}
\end{figure*}

{Figure~\ref{leakandnh} presents the evolution of the $N_{\rm H}$ and 
$f_{\rm leak}$ parameters, alongside direct comparisons of these quantities versus 
$f_{\rm weak}$.}
These results are consistent with our expectation 
that PHL~1811 is subject to Compton-thick obscuration across most epochs.
Specifically,
the observed steep spectral shape arises from the dominance
of the leaked/scattered intrinsic continuum, 
and the extreme X-ray variability is 
driven primarily by variations in the leakage fraction.
The 2024 post-flare observations exhibit some of the largest leakage 
fractions among all weak-state observations, 
and hence PHL~1811 showed the lowest X-ray weakness during these epochs.
Except for the two \swift\ 
observations with $N_{\rm H}<10^{24}$~cm$^{-2}$, 
all weak-state spectra are dominated by the leaked component, 
with only minor contributions from the Compton-reprocessed component. 
The two Compton-thin \swift\ spectra exhibit 
the relatively flat spectral shapes mentioned in 
Section 3. Their lower $N_{\rm H}$ values permit a more 
significant contribution from the transmitted 
component in the hard X-ray band, resulting in the flatter observed spectra.

Our analysis demonstrates that the extreme X-ray 
weakness and variability of PHL~1811 can be 
readily explained by variable obscuration 
from a small-scale, clumpy, dust-free absorber. 
This absorber is likely a clumpy disk wind radiatively 
driven by super-Eddington accretion. 
This result unifies PHL~1811 with the broader 
population of super-Eddington accreting AGNs under 
a single obscuration framework that explains 
their \xray\ and multiwavelength properties.
PHL~1811 probably lies at the high-accretion end of this 
population, hosting a particularly powerful wind 
(e.g., higher density, larger covering solid angle), which
produces its exceptionally large $f_{\rm weak}$ values.
Another object showing similarly large $f_{\rm weak}$ values and variability
amplitudes is PHL~1092,
a PHL~1811 analog at $z=0.396$ \citep{Miniutti2012}.
In terms of variability timescales,
the observed rapid variability of PHL~1811 is not extreme among
the super-Eddington accreting population. For example, \citet{Liu2022}
reported a flux variability factor of $\approx7.6$ within two rest-frame days
for a $z=2.627$ quasar that has a black-hole mass $\approx35$ times larger than that of PHL~1811.

{Recently, we have successfully employed the disk-wind obscuration
scenario to explain the time-resolved X-ray spectra of I~Zwicky~1 (I~Zw~1),
a prototypical NLS1 \citep[][]{Huang2026}.
Its apparent X-ray flares, with 0.3--10 keV count rates varying by a factor 
of $\approx3$ on $\sim10$~ks timescales, 
are explained by reduced obscuration rather than intrinsic coronal flaring.
This work marks an important step toward unifying the remarkable \xray\ properties
of NLS1s and super-Eddington accreting quasars within a single obscuration framework.
A significant number of NLS1s exhibit X-ray weakness and variability
analogous to I Zw 1 \citep[e.g., see Table~6 of][]{Huang2026},
with several objects displaying variability amplitudes comparable to 
those seen in WLQs.
For instance, Mrk~335 shows a maximum X-ray variability
amplitude of $\approx50$ over years,
while its maximum UV variability amplitude
is only $\approx3$, with even milder optical variability
\citep[e.g.,][]{Grupe2007,Komossa2020,Tripathi2020}.
Unlike PHL~1811 and other X-ray weak WLQs,
which often have limited X-ray photon statistics,
these bright NLS1s permit detailed 
X-ray spectral analyses to rigorously test
the obscuration scenario and constrain
the physical properties of the disk-wind absorbers.
Notably, many of these NLS1s (including Mrk 335)
are included in the super-Eddington sample
of \citet{Liu2021},
and their highest X-ray states follow
the canonical $\alpha_{\rm OX}$--$L_{\rm 2500\text{\AA}}$ relation.
This indicates that their variability occurs
between X-ray nominal and \xray\ weak states,
as expected from the obscuration scenario.
Nevertheless, thorough investigations of NLS1 X-ray weakness 
will provide critical insights
into the accretion and outflow physics
of super-Eddington accreting AGNs,
and also probe the dependence of obscuration strength
(wind strength)
on black-hole mass and accretion rate.
}

\section{SUMMARY}

In this paper, we present a systematic analysis of multi-epoch 
X-ray observations of PHL~1811 spanning 2001 to 2024, 
including newly available EP and \swift\ data obtained 
after a 2024 flaring event. Our primary goal is to 
critically examine the two competing physical scenarios 
for its extreme X-ray weakness: heavy line-of-sight 
obscuration and intrinsic X-ray corona suppression. 
We summarize our main findings below:

\begin{enumerate}
\item We confirmed the extreme X-ray weakness 
and strong X-ray variability of PHL~1811 across all 
epochs prior to 2024, with X-ray weakness 
factors $f_{\rm weak}$ ranging from $\approx23$ to $\approx179$. 
We further verified that a simple power-law model fails to 
adequately describe the 2015 \xmm\ spectrum, 
which exhibits a significant hard X-ray excess above $\approx5$ keV. We 
found that the 2006 and 2024 \swift\ observations show 
relatively flat spectral shapes.

\item The 2024 EP WXT detection of a bright flare 
marks the first and only time the source has been 
observed in an X-ray nominal-strength state, 
with $f_{\rm weak}\approx0.63$.
This flare is transient: the X-ray flux declined rapidly after the outburst and partially recovered approximately one month later.

\item We performed spectral fitting for all analyzed observations
using a physically motivated partial-covering obscuration model.
This model provides a good description of the multi-epoch spectra,
with a significant improvement for the 2015 \xmm\ spectrum
relative to the simple power-law model.
The extreme X-ray variability of PHL~1811
is driven primarily by variations in the leakage fraction
and column density of the obscuring material.
The source is subject to Compton-thick obscuration
across most epochs, with the observed steep X-ray spectra
dominated by the leaked/scattered intrinsic continuum.

\end{enumerate}

Our results strongly favor the heavy obscuration scenario over the 
intrinsic X-ray weakness interpretation for PHL~1811.
The variable obscuration can be readily explained by a 
small-scale, clumpy, dust-free absorber 
consistent with a radiatively driven disk wind from 
super-Eddington accretion. This result unifies PHL~1811 
with the broader population of super-Eddington accreting AGNs, 
including WLQs and PHL~1811 analogs, under a single obscuration framework. 

Given the exceptional properties of PHL~1811,
future X-ray monitoring observations are highly worthwhile.
The aims are to capture new X-ray nominal states
and find additional obscuration signatures.
If PHL~1811 is instead detected to be significantly
brighter than the expectation from
the $\alpha_{\rm OX}$--$L_{2500\text{\AA}}$ relation,
the intrinsic corona variation scenario
would be preferred over the obscuration scenario.

~\\

We thank
the referee for carefully reviewing our manuscript and
providing constructive comments.
B.L.\ acknowledges financial support from
the National Natural Science Foundation of China
grant 12573016. W.N.B.\ acknowledges the Penn State Eberly Endowment.
This work is based on data obtained with Einstein Probe,
a space mission supported by the Strategic Priority
Program on Space Science of the Chinese Academy of
Sciences.
We acknowledge the data resources provided by
the China National Astronomical Data Center.
This paper employs a list of 
Chandra datasets, obtained by the 
Chandra X-ray Observatory, contained in~\dataset[DOI:10.25574/cdc.583]{https://doi.org/10.25574/cdc.583}.

%\begin{acknowledgments}

%% Appendix material should be preceded with a single \appendix command.
%% There should be a \section command for each appendix. Mark appendix
%% subsections with the same markup you use in the main body of the paper.
%%
%% Each Appendix (indicated with \section) will be lettered A, B, C, etc.
%% The equation counter will reset when it encounters the \appendix
%% command and will number appendix equations (A1), (A2), etc. The
%% Figure and Table counter will not reset.

\bibliographystyle{aasjournalv7}
\bibliography{refs}

%% This command is needed to show the entire author+affiliation list when
%% the collaboration and author truncation commands are used.  It has to
%% go at the end of the manuscript.
%\allauthors

%% Include this line if you are using the \added, \replaced, \deleted
%% commands to see a summary list of all changes at the end of the article.
%\listofchanges

\end{document}